\documentclass[pre,superscriptaddress,twocolumn,a4paper,showpacs]{revtex4-2}
\usepackage{graphics}
\usepackage{amssymb}
\usepackage{wasysym}
\usepackage{latexsym}
\usepackage{graphicx}
\usepackage{epsfig}
\usepackage{subfigure}
\usepackage{dcolumn}
\usepackage{bm}
\usepackage{comment}
\usepackage{tabularx}
\usepackage{adjustbox}
\usepackage{color, colortbl}
\usepackage{float}
\definecolor{Gray}{gray}{0.9}

\begin{document}

\title{Long-term Scientific Impact Revisited}

\author{Sandro M. Reia}
\affiliation{Instituto de F\'{\i}sica de S\~ao Carlos,
  Universidade de S\~ao Paulo,
  Caixa Postal 369, 13560-970 S\~ao Carlos, S\~ao Paulo, Brazil} 
  
\author{Jos\'e F. Fontanari}
\affiliation{Instituto de F\'{\i}sica de S\~ao Carlos,
  Universidade de S\~ao Paulo,
  Caixa Postal 369, 13560-970 S\~ao Carlos, S\~ao Paulo, Brazil}


\begin{abstract}
Citation based  measures are widely used as  quantitative proxies for  subjective  factors such as the  importance of a paper or even  the  worth of individual researchers. Here we analyze the citation histories  of $4669$ papers published in journals of  the American Physical Society between $1960$ and $1968$ and argue that state-of-the-art  models of citation dynamics and  algorithms for forecasting nonstationary time series are very likely to fail to predict the long-term ($50$ years after publication) citation counts of highly-cited papers using  citation data collected in  a short  period (say, $10$ years) after publication. This is so because those papers do not exhibit  distinctive short-term citation  patterns, although their long-term citation  patterns clearly  set them apart from the other papers. We conclude that even if one accepts that citation counts  are  proxies for the quality of papers, they are not useful  evaluative tools since  the short-term counts are  not informative about the long-term counts in the case of highly-cited papers.
\end{abstract}

\maketitle

\section{Introduction}

``There are things that can be measured. There are things that are worth measuring. But what  can be measured  is not always what is  worth measuring; what gets measured may have no relationship to what we really want to know'' \cite{Muller_2018}.  These words of caution regarding  the indiscriminate use of metrics in today's society make us  wonder whether citations of academic papers  are among those things that are worth measuring.   In our view, the answer is a resounding yes 
when  citations are considered for their own sake  \cite{garfield1979citationBOOK,meho2007rise,fortunato2018science}.  In fact,  citation networks,  citation distributions and citation dynamics are topics that  cover  many of the issues addressed by complexity science  \cite{Phelan_2001}.  In addition and in contrast to  most problems addressed by that novel branch of science, the predictions of the mechanistic models of  citation patterns can  readily be  tested against empirical data available  in citation datasets.

However, when  citation based  measures are used as a  quantitative proxy of a paper's importance \cite{wade1975citation} or as a tool to  evaluate the quality of journals  \cite{garfield1972citation,garfield1979citation} as well as of individual researchers \cite{de2009bibliometrics}, the value of measuring citations is not evident. Here we argue that even if one accepts that  citation counts are a good proxy for those subjective features, they are essentially  worthless as evaluative tools because papers that are highly cited in the long term, say, $50$ years after publication, do not exhibit a distinctive citation  record in, say,  the first $5$  years after publication, which is the typical period used to  evaluate the performance   of individual researchers.  
This very feature makes  the prediction of the long-term citation counts of highly-cited papers using state-of-the-art mechanistic models of citation dynamics \cite{wang2013quantifying} and  algorithms for forecasting nonstationary time series \cite{seabold2010statsmodels} highly inaccurate. 

In particular, in this paper we analyze the citation history of $4669$ papers published in journals of  the American Physical Society (APS) between $1960$ and $1968$ and sort them in 4 classes according to the similarity of the shapes of  their cumulative citation distributions. We find  a strong correlation between the class of a paper and its total citation counts $50$ years after publication.  The citation distribution function  associated to the class that is more likely to include highly-cited papers exhibits a distinctive shape with an inflection point at  about $20$ years after publication that misdirects prediction models trained with data collected in  a short  period (say, $10$ years)  after publication.

The rest of this paper is organized as follows. In  section \ref{sec:dataset} we  characterize the  sample of papers extracted 
from the APS Data Sets for Research \cite{APS_DATA}. Modeling and predicting the citation counts of those papers  are the focus of  the other sections. In particular,  the two mechanistic models of citation dynamics  used to explain the citation counts, viz., the WSB model \cite{wang2013quantifying} and the SIR epidemic model \cite{reia2021sir}, are   described briefly in  section \ref{sec:models} together with the basic ARIMA (AutoRegressive Integrated Moving Average) model \cite{seabold2010statsmodels}  for time series prediction. The  fitting and  prediction performances of those models are then discussed in sections 
\ref{models_goodness} and \ref{predicting_citations}, respectively. Finally, section \ref{sec:disc} is reserved to our concluding remarks.

\section{The APS Dataset}
\label{sec:dataset}

The APS Data Sets for Research \cite{APS_DATA} comprises citing article pairs and bibliographic meta-data of papers published in the APS journals since $1893$. Here  we focus   on the sample of $4669$ papers published between $1960$ and $1968$ that received at least $10$ citations during the five years period  after  their publication, which amounts to $20.75\%$ of the papers published in  the APS journals in that period. The selected sample comprises $2975$ papers published in  \textit{Physical Review}, $1586$ in  \textit{Physical Review Letters} and $108$ in \textit{Reviews of Modern Physics}. We track the number of citations $c_i(t)$ that paper $i = 1, \ldots, 4669 $ in this sample  received up to $t \leq 50$ years after  its publication.

\begin{figure}[t!]
\centering  
\includegraphics{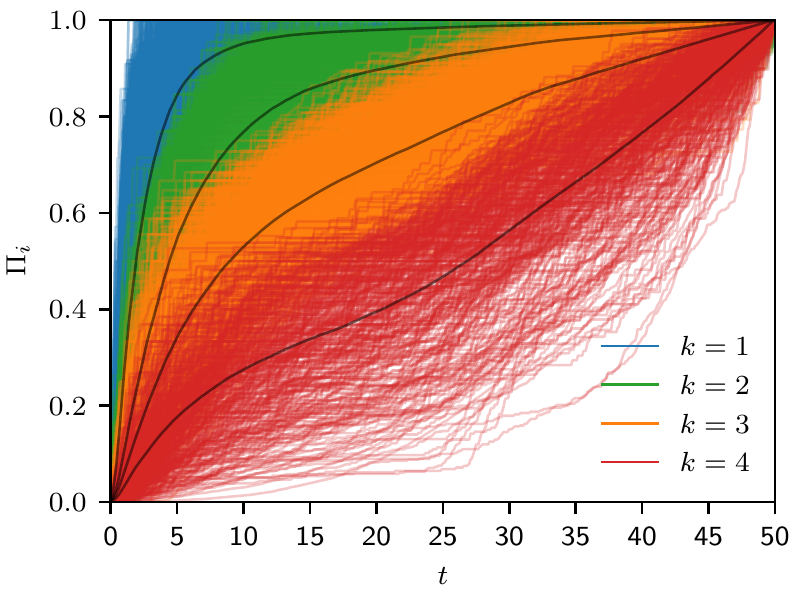}
\caption{Cumulative distributions  of citations $\Pi_i$ received by each targeted paper  as function of the number of years  after publication. The cumulative curves of the $4669$ targeted  papers are clustered into $K = 4$ classes. There are $1686$ papers in class $k = 1$, $1654$ in class $k = 2$, $955$ in class $k = 3$ and $374$ in class $k = 4$. The black curves  indicate the average cumulative distributions  for  each class.}
\label{fig1}
\end{figure}

\begin{figure*}[htb!]
\centering  
\includegraphics{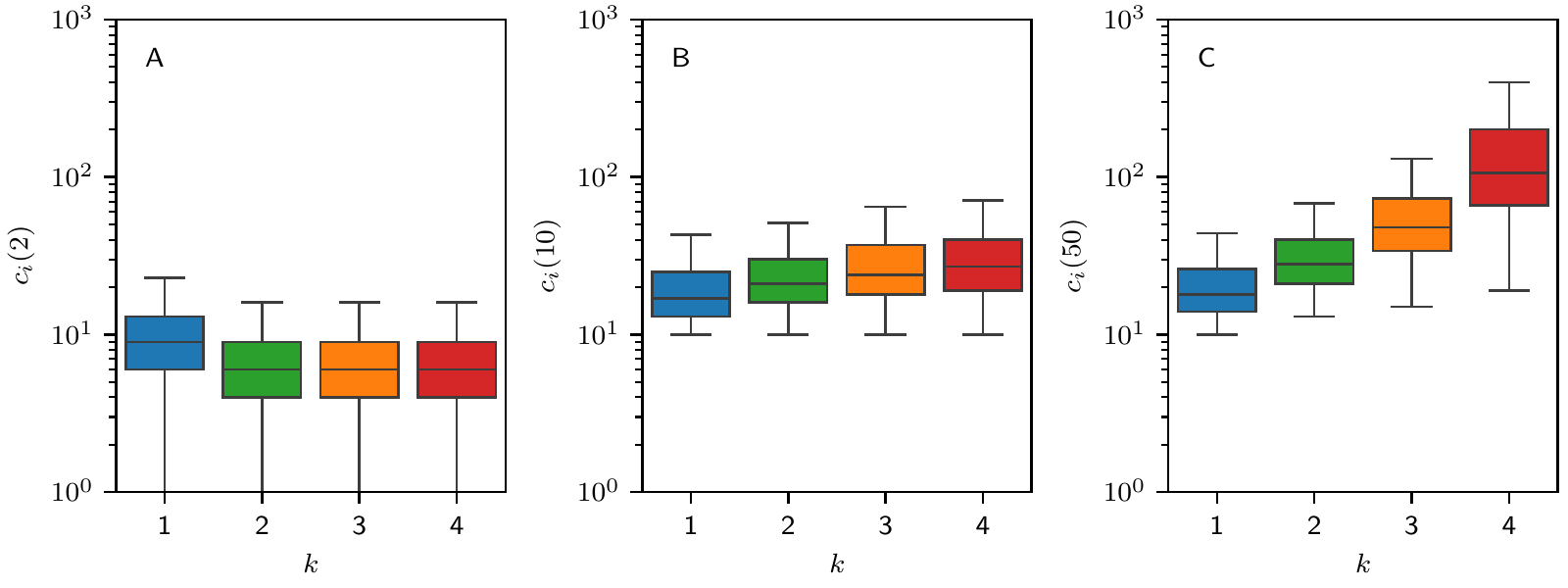}
\caption{Boxplots of the number of citations received by  papers in classes $k=1,2,3,4$ at  $t=2$ years (panel A),  $t=10$ years (panel B)  and  $t=50$ years (panel C) after  publication.}
\label{fig2}
\end{figure*}

The cumulative distribution functions $\Pi_i (t)  = c_i(t)/c_i(50)$ of the citations received by each   targeted paper  in the $50$ years period considered  are shown in Fig. \ref{fig1}, 
where we have used the $K$-means clustering algorithm \cite{steinley2006k} to sort those papers into $K = 4$ classes according to the similarity of their cumulative distributions.  The black curves in this figure indicate the typical (or average) cumulative distribution of each class. Of particular interest are the citations received in the  first two years after publication  since this is the period used to measure the  Impact Factor (IF)  of a journal \cite{Garfield_2006}: papers in class $k = 1$ received more than $50\%$ of their citations  in that period, papers in  class $k = 2$ received about $24\%$, papers in class $k = 3$, $14\%$  and papers in class $k = 4$ received a meager $7\%$. As illustrated in Fig. \ref{fig2}, which shows the  boxplots of the number of citations received  by papers in class $k$ at  $t$ years after publication,  there are no significant differences in the  distribution of citations received by  papers of distinct classes in that short time span (i.e., for $t=2$ years).

The boxplots of  Fig. \ref{fig2}  reveal the intriguing finding  that papers in class $k = 4$ receive, on the average, more citations than the papers in the other classes for sufficiently large time spans. In fact, the average (or, more precisely, the median) number of citations increases monotonically with the class index $k$ in panels B and C that show the citation counts for  $t=10$  and $t=50$ years after publication, respectively.    Hence, papers in class $k=4$ are likely to be among the most cited ones.  In fact,  the odds that a randomly selected paper in class $k=4$  is among the $10\%$ most cited papers in our sample of $4669$   papers is about $60 \%$. We emphasize that  these results are not  consequences of the definition of classes,  which are determined  by the similarity of the shapes of the citation patterns in Fig.  \ref{fig1} and  do not use  information about  the number of citations received by the targeted papers.

Since  the number of classes $K$ is an input parameter to the  $K$-means clustering algorithm, a word is in order about the choice $K=4$.  Of course,  we have tried  many different choices of $K$, each choice resulting in a variant of Fig.  \ref{fig1}.  On the one hand, we have found   it  difficult to spot qualitative  differences between the   average cumulative distributions associated to adjacent  classes for $K> 4$.  On the other hand,  for $K < 4$ the  average cumulative distribution that exhibits an  inflection point and that characterizes class $k=4$ disappears because of the merging of classes $k=3$ and $k=4$. Hence the choice $K=4$.

\section{Models of citation dynamics}\label{sec:models}

Here we focus  on two mechanistic models of citation dynamics, viz.,  the WSB model and the SIR epidemic model,   that can explain a large variety of citation patterns by tunning  a few parameters only.  In this section  we offer a brief account of these two models and in the next sections we compare their fitting and  prediction performances  with the basic ARIMA (AutoRegressive Integrated Moving Average) model \cite{seabold2010statsmodels}, which is widely used in  fitting and forecasting nonstationary time series  \cite{chatfield2000time,hyndman2018forecasting}. Since  the ARIMA model is not a model of citation dynamics, we will not examine it any further in this paper. 

The aim of the mechanistic models, as well as of ARIMA, is to fit the  empirical  total citation counts  $c_i (t) $ received by paper $i$ up to  $t$ years after its publication. In order to distinguish between the empirical citation counts $c_i(t)$  and the citation counts predicted by the models we use the notation  $\tilde{c}_i (t) $ for the latter.

\subsection{The WSB model}\label{model_wsb}

The WSB model is a  successful mechanistic model of citation dynamics  that builds on three assumptions, viz., preferential attachment,  fitness and  aging    \cite{wang2013quantifying}.  The name of the model is an acronym for the name of their proponents. 
Preferential attachment  means that the probability
that a publication is cited is an increasing function of its current number of citations \cite{Price_1975,Merton_1973,Redner_2005}. Fitness expresses the notion that papers differ with respect to the perceived novelty and importance of their contents \cite{Foster_2015,Li_2019}, whereas aging captures the fact  that the perceived novelty and importance of a paper  eventually  fade out \cite{Eom_2011}.  Although there are  many intangible factors behind an author's decision to cite a paper, such as the  reputation of its authors and of the  journal where it was published  \cite{Dong_2016} that cannot be  described by  a mechanistic model, the WSB model does a remarkably 
good job at predicting  long-term citations of papers in classes $k=1$ and $k=2$ as we will show in this paper.

More pointedly, the  WSB model expresses  the total citations counts received by paper $i$ up to time $t$ after its publication through a disarmingly simple formula  \cite{wang2013quantifying}
\begin{equation}\label{eq_wsb}
	\tilde{c}_i(t) =  m \left[e^{\lambda_i \Phi\left( \frac{\ln t - \mu_i}{\sigma_i} \right)} - 1 \right] ,
\end{equation}
where $\Phi(x) =  \int_{-\infty}^{x} e^{-y^2/2} dy/\sqrt{2 \pi}$. Here the paper-dependent parameters $\lambda_i$, $\mu_i$ and $\sigma_i$ are  related to  the relative importance of paper $i$ with respect to the other papers,  the time taken for paper $i$ to reach  its citation peak, and  its longevity, respectively. Those parameters are obtained by fitting  eq. (\ref{eq_wsb}) to the data describing the citation history of paper $i$. The parameter $m$ is the mean number of references of the papers in the  sample considered, which  has little effect on the overall predictive performance of the model and so it is set  to the fixed value $m = 30$ for all papers in the sample \cite{wang2013quantifying}.



\subsection{The SIR epidemic model}\label{model_sir}

In line with  the seminal attempt to describe the spread  of Feynman diagrams through the theoretical physics communities of different countries using models of epidemics  \cite{bettencourt2006power}, the SIR epidemic  model \cite{kermack1927contribution} was used  recently to fit the citation history  of  highly-cited papers \cite{reia2021sir}. 
In this framework, the total number of citations paper $i$ receives up to time $t$  is 
\begin{equation}\label{eq_upsilon}
	\tilde{c}_i(t) = S_i (0) - S_i(t) ,
\end{equation}
where $S_i(0)$ is the number of papers in an abstract population of  papers not yet written that are susceptible to cite paper $i$ and  $S_i (t) \leq S_i(0)$ is the  number of  citations  paper $i$  can still receive after time $t$. The decrease in the number of susceptible papers is determined by a contact process that mimics the spread of an infectious disease, i.e., 
\begin{equation}
	\frac{dS_i}{dt} = -\beta_i S_i (t) \frac{I_i (t)}{N_i}, 
	\label{eq_susceptible}
\end{equation}
where $I(t)$ is the number of papers that have cited  paper $i$  before or at  time $t$ and that  can still influence susceptible papers  to cite that paper. The parameter $\beta_i$  is a measure of the degree of persuasion of the influential papers and $N_i = S_i(0) + I_i(0)$ is a constant.  The equation for the number of influential papers is
\begin{equation} \label{eq_infected}
	\frac{dI_i}{dt} = \beta_i S_i (t) \frac{I_i (t)}{N_i} - \gamma_i I_i (t), 
\end{equation}
where the parameter $\gamma_i$ controls the rate at which the influential papers cease to be persuasive  to produce new citations of paper $i$. Hence,  the SIR epidemic model has three free parameters, viz.,  $S_i(0)$, $\beta_i$ and $\gamma_i$ that must be tuned to fit the empirical cumulative number of citations of   paper $i$.  

We find that  the SIR epidemic model exhibits a fitting and prediction performances practically indistinguishable from those of the Bass model  that builds on the mechanisms that drive the adoption  of a new product (viz.,  innovation and imitation)   to explain the citation dynamics \cite{Mingers_2008,Min_2018}. This is the reason  we will not consider the Bass model in this contribution.

\begin{figure*}[htb!]
\centering  
\includegraphics{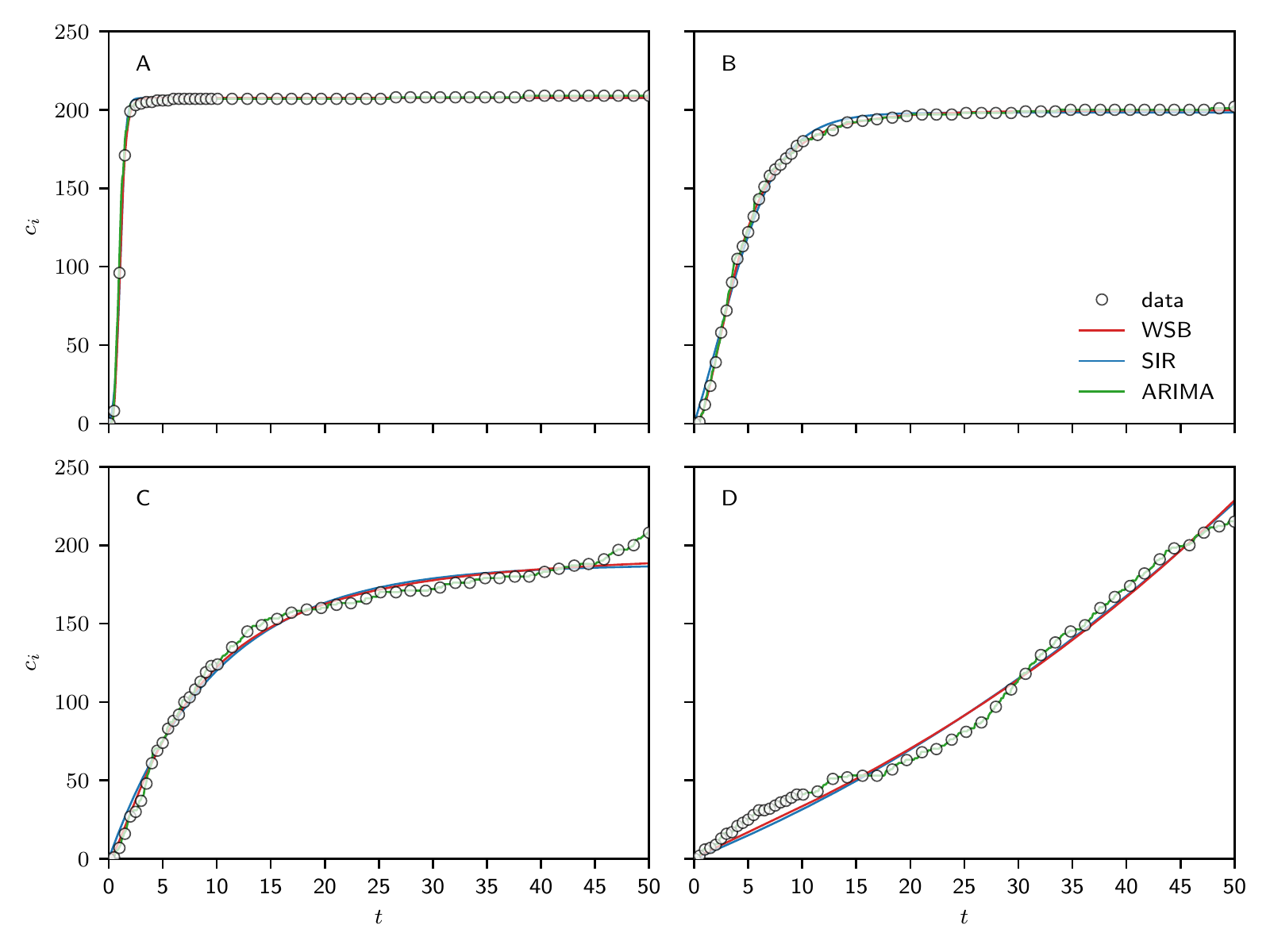}
\caption{Citation counts $c_i $ as function of time $t$  in years for representative  papers in classes $k = 1$  (panel A), $k = 2$ (panel B),  $k = 3$ (panel C),  and $k = 4$ (panel D). The curves are the fittings produced by the WSB, SIR and  ARIMA models as indicated. The curves for the two mechanistic models are practically indistinguishable  whereas the ARIMA model fits the empirical citation counts  perfectly. }
\label{fig3}
\end{figure*}

\section{Goodness of fit}\label{models_goodness}

Here we address the fitting performances of the models  discussed in the previous section.  Figure \ref{fig3} illustrates the citation histories of four papers  that are representative of  the different classes and that  have approximately the same number of  citations (about $210$) in the 50 years period considered.  In particular, panel A  shows the citation history of paper  \cite{rosenfeld1967data}  in class $k=1$, panel B of paper \cite{PhysRev.138.B913} in class $k=2$,  panel C of paper  \cite{perey1963deuteron} in class $k=3$, and panel D of paper  \cite{PhysRevLett.10.516} in class $k=4$. The symbols in this figure are the empirical citation counts $c_i(t)$ and the curves are the citation counts produced by the models $\tilde{c}_i(t)$.
Regardless of the class considered, there are no perceivable differences between the fittings produced by the WSB and SIR models and both models exhibit a somewhat wanting  performance for the papers in classes $k=3$ and $k=4$. As expected,  the ARIMA model fits the data perfectly since,  unlike the mechanistic models,  it is not constrained by a fixed  functional form. 

\begin{figure*}[htb!]
\centering  
\includegraphics{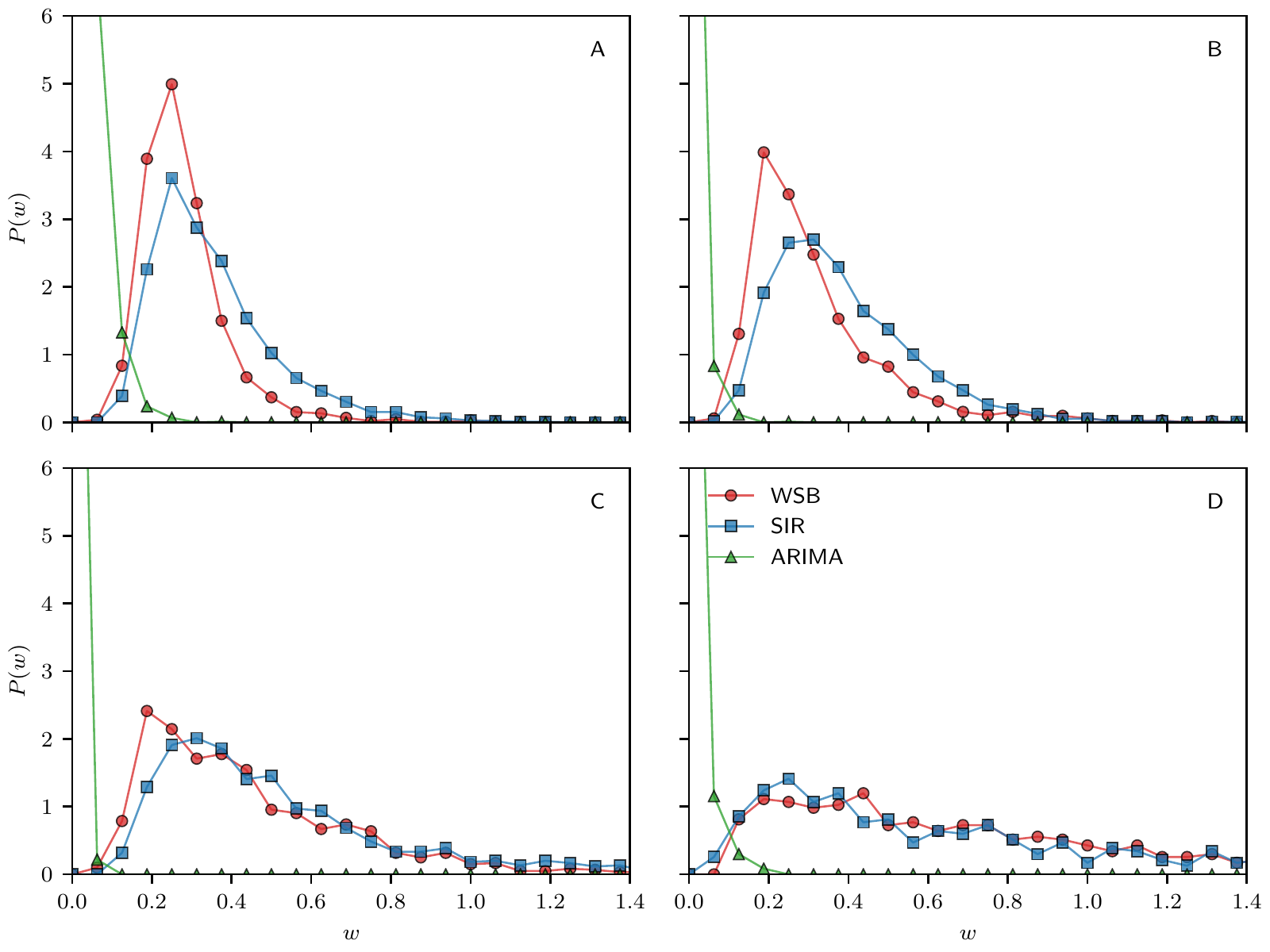}
\caption{Distribution of probability $P(w)$ of the  weighted KS measure [see eq.\ (\ref{eq_wks})] of papers in  classes $k = 1$ (panel A), $k = 2$  (panel B), $k = 3$ (panel C) and $k = 4$ (panel D) for  the  WSB, SIR  and ARIMA models as indicated.}
\label{fig4}
\end{figure*}

\begin{figure*}[htb!]
\centering  
\includegraphics{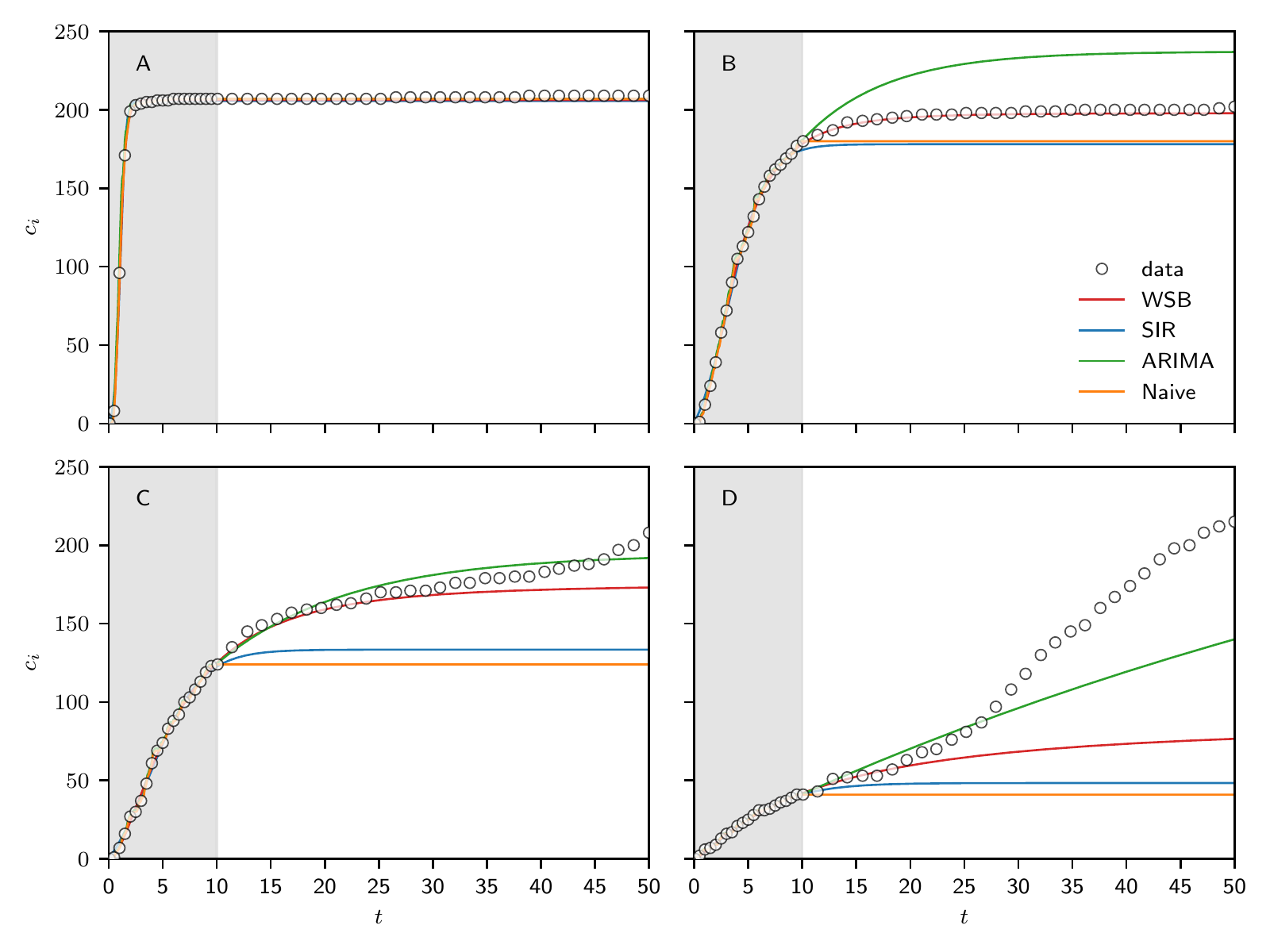}
\caption{Citation counts $c_i$ as function of time $t$ in years for representative  papers in classes $k = 1$  (panel A), $k = 2$ (panel B),  $k = 3$ (panel C),  and $k = 4$ (panel D). These are the same papers of Fig.\ \ref{fig3}.  The curves are the predictions of the  WSB, SIR, ARIMA and naive expectation models, as indicated, with the parameters adjusted to fit the empirical citation counts in the range $t \in [0,10]$ years (gray region). 
}
\label{fig5}
\end{figure*}

The analysis of the results exhibited in Fig.\ \ref{fig3} offers only a qualitative assessment of the goodness of fit of the models for  particular papers. A more useful and robust quantitative measure is  the weighted KS test \cite{wang2013quantifying} given by
\begin{equation}\label{eq_wks}
	w_i = \max_{t \in [0, T]} \frac{|c_i(t) - \tilde{c}_i(t)|}{\sqrt{[1 + c_i(t)][c_i(T) - c_i(t) + 1]}},
\end{equation}
where $T$ is the upper limit  of the fitting range, which in our case is $T=50$ years. Essentially, $w_i$ picks the largest deviation between the theoretical and empirical counts for paper $i$  in the  entire  fitting range, so it is a worst-case measure. Since for each paper we have a value of $w_i$ we can consider the distribution of probability $P(w)$  for  papers in the different classes, which is shown in Fig. \ref{fig4}. A good fitting performance is signaled by a high peak of $P(w)$ at very low values of $w$, as exhibited by  the ARIMA model.  The quality of the fitting decreases as the right-tail of the distribution $P(w)$  increases.   The weighted KS measure allows us to realize that the WSB model marginally outperforms  the SIR model for all classes except for class $k=4$ for which both models exhibit a similar very poor performance. Hence, our findings show that the mechanistic models considered fail to explain the long-term citation histories of  highly-cited papers, which prompts the   problem of   how  to modify the WSB model   in order  to fit the citation history of papers in class $k=4$.  We will not address this attractive issue in this contribution, however.

Of course,   fitting citation counts makes sense only for the mechanistic models since the interpretation of the model parameters can yield valuable information about the characteristics of  the targeted papers such as their  perceived novelty and importance \cite{wang2013quantifying,reia2021sir}. In contrast, we learn nothing by fitting  citation histories  with the ARIMA model since its parameters are not interpretable  in terms of the citation dynamics. Nevertheless,  we choose to consider the ARIMA model in this section because it   illustrates nicely  the distribution $P(w)$  for   models that  fit the data very well. The practical use of the ARIMA model is the prediction of citation counts that we address in the next section.

\section{Predicting citation counts}\label{predicting_citations}

Behind the study of  mechanistic models of citation dynamics is, of course, the issue of whether the citation counts of a particular paper can be predicted or not, which is the topic of this section.  In addition to the  three models used  in the previous section to fit the citation histories of papers, in this section we consider  the naive expectation of citation counts, which assumes  that  papers do not receive new citations after the training period. The  naive expectation  plays the role of  a null model to assess the quality of  the predictions of the  WSB, SIR and ARIMA models \cite{wang2014science}.

To predict the citation counts of a paper, the parameters of the models are tuned to fit the empirical citation counts in a certain training period. Here, we set the training period  to $10$ years after  publication.
Figure \ref{fig5} shows the prediction performances of the four models for the same papers exhibited in Fig. \ref{fig3}. We observe that  all models fit the empirical citation counts  very well  in the training period, which is highlighted by the gray background in the figure.  The models predict  accurately the citation counts of the representative  paper of class $k=1$, but this  success is obscured by the fact that the prediction of the naive expectation model is  equally accurate. In fact, we will show later that the naive expectation is the best predictor for papers in class $k=1$, which amounts to  $36.1\%$ of the papers in our sample.  The  real challenge is predicting citation counts of papers in the other classes and  Fig. \ref{fig5} indicates that the WSB model is the most consistent predictor of all the models considered, in the sense that it predicts accurately the long-term citation counts of the representative paper of class $k=2$,  the middle-term citation counts of the  representative paper of class $k=3$ and the  short-term citation counts of the representative paper of class $k=4$. The prediction performance of the SIR model is comparable to the performance of the null model for the four papers considered.  The disastrous long-term prediction performances of all models for the  representative paper  of class $k=4$ is another  evidence that the mechanistic models considered here  are not suitable to describe the citation history of papers in that class.

\begin{figure*}[htb!]
\centering  
\includegraphics{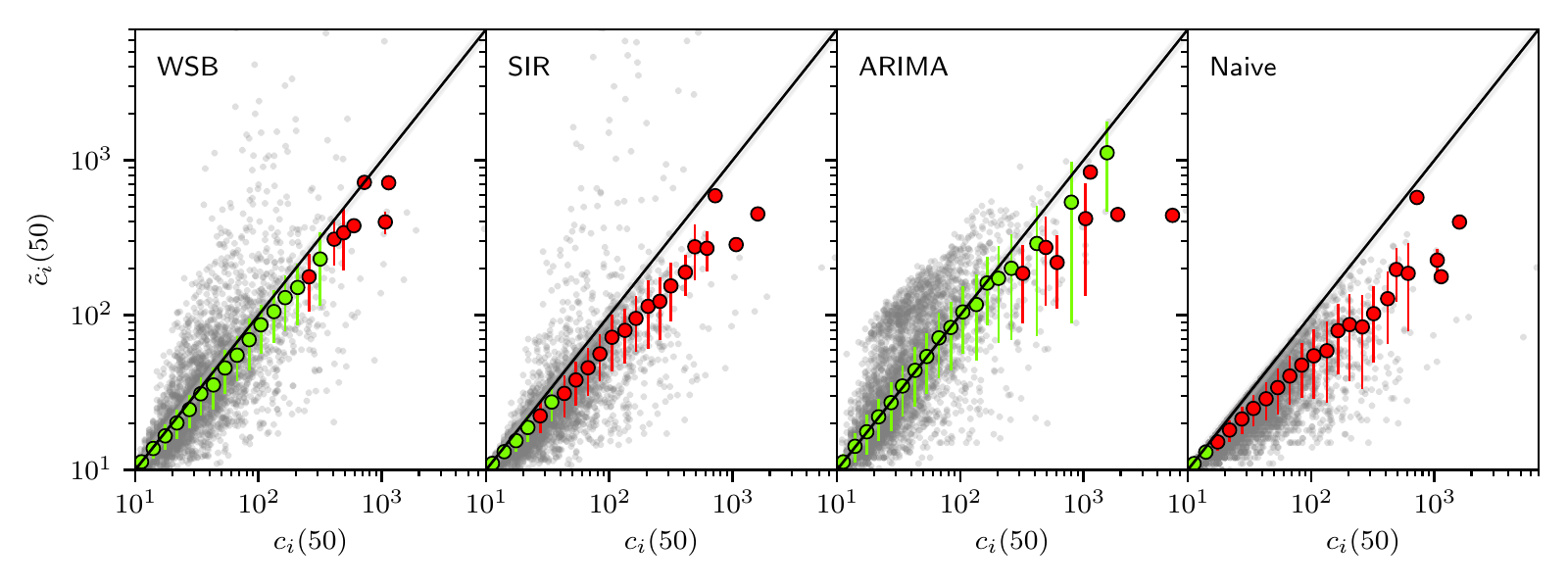}
\caption{Scatter plots of the predicted and real citation counts at  $t=50$ years after publication. The model parameters were adjusted in a training period of $10$ years after publication. Each gray symbol corresponds to  a paper and the colored symbols  represent the average of the predictions within the corresponding bin. Green symbols indicate that the mean predicted citation counts are within one standard deviation from the actual counts, whereas the red symbols indicate that those counts are more than one standard deviation apart from the actual counts.
}
\label{fig6}
\end{figure*}

Figure \ref{fig6} exhibits a qualitative method to assess the long-term predictive power of the models for all the $4669$ papers in our sample. The scatter plot for each model shows  the real number of citations of a paper at $t=50$ years after publication (i.e., $c_i(50)$)  in the x-axis and the predicted number  (i.e., $\tilde{c}_i(50)$) in the y-axis. We recall that the model parameters were adjusted in the time window $t \in \left [ 0, 10 \right  ]$ years.  Since  all  gray points (each point  correspond to a paper) should lie on the diagonal for the  perfect predictor,  the distance  of the points to the diagonal is a indication of the quality of the prediction. In fact, collecting the papers which have a similar number of citations in a same bin allows us to estimate the mean and the standard deviation of the predicted counts for each bin. These mean citation counts  are represented by the colored symbols in the figure: if  the mean is less than one standard deviation apart from the diagonal the symbol is colored green, otherwise  it  is colored red.  The density of point above (below) the diagonal measures the tendency of  the models to overestimate (underestimate) the citation counts.  We note that  the naive expectation model always underestimates the citation counts.
The results are consistent with our findings for the representative papers of each class summarized in Fig.\ \ref{fig3}: the WSB and the ARIMA models are the best predictors and their average prediction  performances are poor for highly-cited (i.e., class $k=4$) papers only.

\begin{figure*}[htb!]
\centering  
\includegraphics{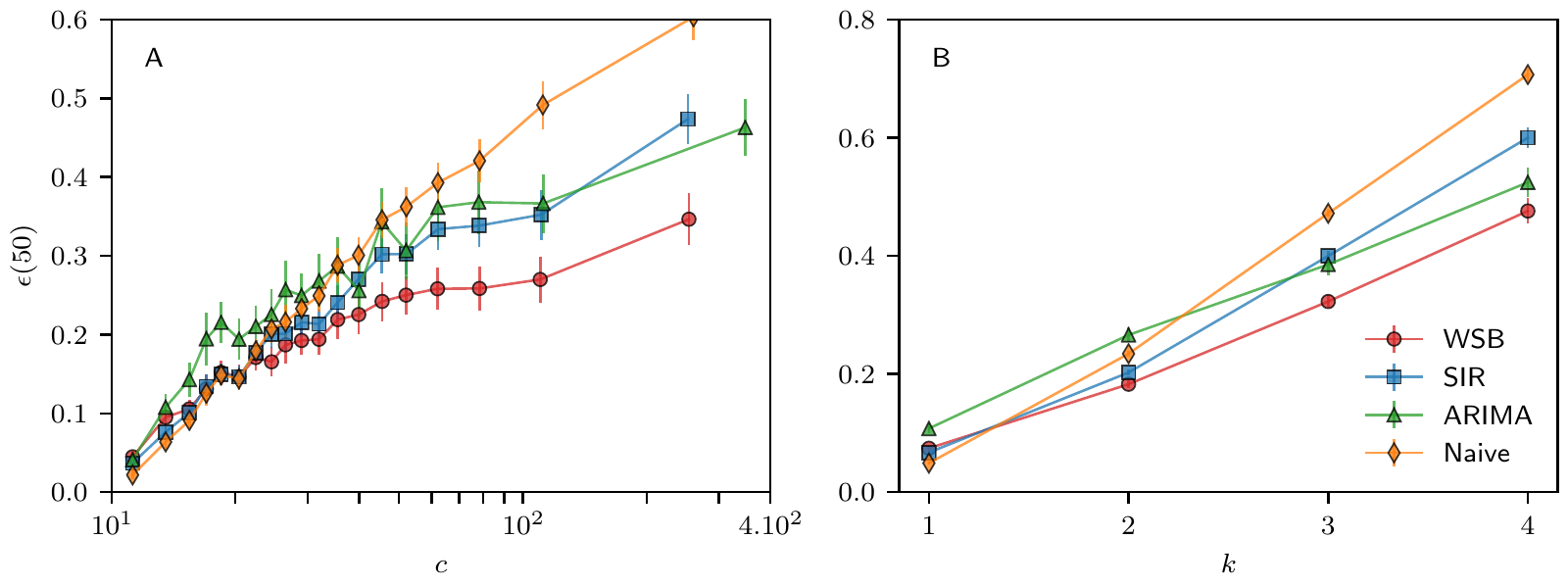}
\caption{Mean absolute percentage error $\epsilon$ at $t=50$ years as function of  the number of citations $c$ (panel A) and of the paper class $k$ (panel B)   for the    WSB, SIR, ARIMA and naive expectation models, as indicated. The model parameters were adjusted in a training period of $10$ years after publication. }
\label{fig7}
\end{figure*}

In order to extract  quantitative information from the scatter plots of  Fig. \ref{fig6}, we consider the mean absolute percentage error $\epsilon$ defined as  \cite{shen2014modeling} 
\begin{equation}\label{eq_mape}
	\epsilon (t)  = \frac{1}{|  \Omega | } \sum_{i \in \Omega} \left| \frac{c_i(t) - \tilde{c}_i(t)}{c_i(t)} \right| ,
\end{equation}
where $\Omega$ is  a set of interest (e.g., the set of papers in class $k$)  and $|  \Omega |$ stands for  the  cardinality of $\Omega$, as usual. The lower the  value of $\epsilon (t)$, the better the  prediction performance  at time $t$ for  papers  in  the set $\Omega$.  Figure \ref{fig7} shows the  mean absolute percentage error at $t=50$ years after publication for the set of papers whose citation counts fall in the bin centered at $c$ (panel A) and for the set of papers in class $k$ (panel B).  Of course, since the class of a paper is strongly correlated with its  citation count, these two panels yield essentially the same information. For all models,  the mean absolute percentage error  increases with the number of citations in accordance with our qualitative analysis   of Fig.\ \ref{fig6}. As already pointed out, the naive expectation model exhibits the best prediction performance for papers in class $k=1$   and, somewhat surprisingly,  the ARIMA  model exhibits the worst performance for papers in class $k=1$ and $k=2$.   The WSB model significantly outperforms the other models for  classes  $k=2, 3$ and $4$. However, the mean absolute percentage error of about $40\%$ for papers in class $k=4$ reinforces the sense of inadequacy of the mechanistic models to describe and predict the long-term citation counts of highly-cited papers.

Since the prediction performance of the WSB model has already been addressed  in the literature \cite{wang2013quantifying,shen2014modeling,wang2014science},  it is appropriate to highlight  our original contributions to this issue,  which are twofold. First,  the previous studies considered the citation histories and citation predictions up to $t=30$ years after publication, whereas  here we have extended that range to  $t=50$ years. Although this extension  makes no difference  for papers in classes $k=1$ and $k=2$, it is necessary to expose the inadequacy of the WSB model to describe papers in classes $k=3$ and $k=4$ (see Fig.\ \ref{fig5}). Second and most importantly, we have used the K-means clustering algorithm  to sort the papers in $K=4$  classes according to the shape of their citation cumulative distributions (see Fig.\ \ref{fig1}). Somewhat surprisingly, we have found that those classes correlate strongly with the citation counts of the papers and that the  prediction performance of the WSB model, as well as of the other models considered here, varies  greatly depending on the class of the target paper (see Fig. \ref{fig7}). In particular, the  WSB model does a remarkably good job at predicting the long-term citation counts of papers in classes $k=1$ and $k=2$, which comprise 
$71.5\%$ of the papers in our sample, but  fails unarguably for papers in class $k=4$, which comprises  $8\%$ of the papers  only. We recall that papers in class $k=4$,  however, are very  likely to be among  the most cited papers in our sample.

\section{Conclusion}\label{sec:disc}

Our results imply that use of citation based  measures collected in a relatively short  period (typically five years after publication)  as a  quantitative proxy of a paper's importance is unfounded, even if one accepts that the number of citations correlates strongly with the (subjective) notion of  the importance of a scientific contribution. This is so because papers that are highly cited in the long term (say, $50$ years after publication) do not exhibit a distinctive citation  record in the first years after publication (see Fig.\ \ref{fig2}). This feature makes the prediction of the long-term citation counts of those papers using known mechanistic models of citation dynamics  and  algorithms for forecasting nonstationary time series basically useless (see Fig.\ \ref{fig7}). However, once we know the entire citation history of a paper we can realize that  highly-cited papers exhibit a very distinctive citation pattern, which is easily singled out by  the K-means clustering algorithm  (see Fig.\ \ref{fig1}).  Unfortunately, this sort of  information has no predictive value since, as pointed out,  to draw the cumulative distribution of citations we must know the complete citation counts of a paper.

\section*{Acknowledgments}
We thank the American Physical Society for letting us use their citation database.
The research of JFF was  supported in part 
 by Grant No.\  2020/03041-3, Fun\-da\-\c{c}\~ao de Amparo \`a Pesquisa do Estado de S\~ao Paulo 
(FAPESP) and  by Grant No.\ 305058/2017-7, Conselho Nacional de Desenvolvimento 
Cient\'{\i}\-fi\-co e Tecnol\'ogico (CNPq).
SMR was supported by the Coordena\c{c}\~ao de Aperfei\c{c}oamento de Pessoal de
N\'{\i}vel Superior - Brasil (CAPES) - Finance Code 001.

%



\end{document}